\documentstyle[aps,12pt]{revtex}

\begin{document}
\draft
\author{O. B. Zaslavskii}
\address{Department of Mechanics and Mathematics, Kharkov V.N. Karazin's\\
National\\
University, Svoboda\\
Sq.4, Kharkov 61077, Ukraine\\
E-mail: ozaslav@kharkov.ua}
\title{Classical model of elementary particle with Bertotti-Robinson core
and extremal black holes}
\maketitle

\begin{abstract}
We discuss the question, whether the Reissner-Nordstr\"{o}m (RN) metric can
be glued to another solutions of Einstein-Maxwell equations in such a way
that (i) the singularity at $r=0$ typical of the RN metric is removed (ii),
matching is smooth. Such a construction could be viewed as a classical model
of an elementary particle balanced by its own forces without support by an
external agent. One choice is the Minkowski interior that goes back to the
old Vilenkin and Fomin's idea who claimed that in this case the bare
delta-like stresses at the horizon vanish if the RN metric is extremal.
However, the relevant entity here is the integral of these stresses over the
proper distance which is infinite in the extremal case. As a result of the
competition of these two factors, the Lanczos tensor does not vanish and the
extremal RN cannot be glued to the Minkowski metric smoothly, so the
elementary-particle model as a ball empty inside fails. We examine the
alternative possibility for the extremal RN metric - gluing to the
Bertotti-Robinson (BR) metric. For a surface placed outside the horizon
there always exist bare stresses but their amplitude goes to zero as the
radius of the shell approaches that of the horizon. This limit realizes the
Wheeler idea of ''mass without mass'' and ''charge without charge''. We
generalize the model to the extremal Kerr-Newman metric glued to the
rotating analog of the BR metric.
\end{abstract}



\section{Introduction}

Among other remarkable things, a black hole is sometimes considered as a
classical model (or analogue) of an elementary particle \cite{mar73}, \cite%
{wil}. For an external observer it reveals itself like an object with a few
parameters such as a mass $m$ or charge $e$, while it also can contain a
rich structure inside. In this context there is temptation to remove
singularities, typical of inner region of black holes. In particular, the
Schwarzschild singularity inside may be replaced by the de Sitter (dS)
region, if the hypothesis about the limiting curvature is accepted but this
demands some transition layers with delta-like stresses inside the
Schwarzschild region \cite{fmm}, \cite{many}. Similar stresses appear on the
horizon itself \cite{gh} if the whole inner region of the Schwarzschild
region is replaced by the dS one \cite{gd}. Another possibility is to
consider matter with the equation of state $p_{r}=-\varepsilon $ ($p_{r}$ is
the radial pressure, $\varepsilon $ is the energy density) that can lead to
regular black holes with the dS core \cite{dym1}, \cite{dym2}.

In the case of a charged black hole, there exist the following variants.
First, the solution can represent an usual RN black hole that, as is well
known, contains a singularity (hidden beyond the horizon, if $m\geq e$).
Second, if sources are distributed on the sphere, one can consider matching
the RN metric outside with the Minkowski region inside. In doing so, the
region with a singularity inside a horizon is removed and replaced by an
empty Minkowski spacetimes. As far as the problem of the self-energy is
concerned, the event horizon manifests itself for an external observer as a
regulator due to gravitational effects, removing divergencies typical of a
point particle in classical electrodynamics. If bare stresses on the shell
vanish, one would obtain the pure field model of a ''classical electron'' in
the spirit of Abraham and Lorentz. This idea found explicit realization in
the paper by Vilenkin and Fomin \cite{vf} who claimed that such a field
model is self-consistent for the extremal case $m=e$ only.

However, more thourough analysis presented below does not confirm the
conclusion about smooth matching between extremal RN and Minkowski regions.
In the coordinate frame where all metric coefficients are continuous, the
amplitude of the delta-like stress-energy tensor of the extremal
configuration does tend to zero on the horizon. However, in spite of this,
the Lanczos tensor (obtained by integration over the proper distance across
the shell) does not vanish because of an infinite proper disctance. It was
stated in \cite{vf} that if the black hole is extremal ($m=e$), one can
define the energy-momentum vector entirely in the outer region, whereas the
region beyond the horizon contributes nothing into dynamical characteristics
in any frame. We demonstrate, however, that this conclusion of \cite{vf} is
not covariant: it is valid in static coordinates but are violated in the
frame of a free-falling observer.

Meanwhile, in the case $m=e$ there is also the third possibility - the
so-called Bertotti-Robinson (BR) solution which is not
spherically-symmetrical since it does not possess a center at all and is
regular everywhere \cite{r}, \cite{b}. The aim of the present paper is to
examine the corresponding possibility to construct a self-consistent model
of a \textquotedblright classical electron\textquotedblright\ by gluing the
external extremal RN metric to the BR one inside. Such a construction is (i)
free from singularities, (ii) self-supporting in the sense that there is no
external agent to maintain it in the equilibrium, (iii) asymptotically flat.
However, the important reservation is in order. Both properties (i) and (ii)
should be understood in the sense of the limiting transition only. For a
shell placed at the $r_{0}=r_{h}+\varepsilon $ ($r_{h}$ is a horizon radius)
there always exist singular bare stresses on the shell but their amplitude
tends to zero as $\varepsilon \rightarrow 0$. In other words, the
construction under consideration represents a regular limit of singular
configurations. Instead of a sophisticated structure inside a black hole our
construction is in a sense as simple as possible there, being a direct
product {\rm AdS}$_{2}\times S^{2}$. In this sense it can be viewed as an
alternative to the black holes with the dS core \cite{dym1}, \cite{dym2}.

We discuss the energy contribution from the horizon and, as a by-product, we
establish some features of quasilocal energy-momentum \cite{by}, \cite{bly}
inherent to the generic (not necessarily extremal) spherically-symmetrical
horizon. We also generalize our construction to the Kerr-Newman black hole.

\section{Basic equations}

Let us consider the metric of a static spherically symmetric black hole
spacetime

\begin{equation}
ds^{2}=-b^{2}dt^{2}+dl^{2}+r^{2}(l)d\omega ^{2}\text{.}  \label{b}
\end{equation}

We assume that there are two regions to be glued together along $r=r_{0}$.
We denote these regions as ''+'' for $r\geq r_{0}>r_{h}$ and ''-'' for $%
r\leq r_{0}$, where $r_{h}$ corresponds to the horizon.

We would like the metric to be continuous across the shell but, in general,
the terms with first derivatives may acquire jumps. It means that there is
some effective transition layer at $r=r_{0}$. If we write down the Einstein
equations for the system with such a layer in the form

\begin{equation}
G_{\mu }^{\nu }=8\pi (T_{\mu }^{\nu }+\tilde{T}_{\mu }^{\nu })\text{,}
\end{equation}
where $\tilde{T}_{\mu }^{\nu }$ is the delta-like contribution from the
layer, the conditions of smooth matching read \cite{k}

\begin{equation}
S_{\mu }^{\nu }\equiv \int_{r_{0}-0}^{r_{0}+0}dl\tilde{T}_{\mu }^{\nu }=0%
\text{,}
\end{equation}%
where $S_{\mu }^{\nu }$ is the so-called Lanczos tensor. Following the
general formalism \cite{k}, one can write 
\begin{equation}
8\pi S_{\mu }^{\nu }=[K_{\mu }^{\nu }]-\delta _{\mu }^{\nu }[K]\text{,}
\label{s}
\end{equation}%
where $K_{\mu }^{\nu }$ is the tensor of the extrinsic curvature, $%
K=K_{i}^{i}$ ($i=0,2,3$) and $[...]=(...)_{+}-(...)_{-}$. If $[K_{\mu }^{\nu
}]=0$, then both region match smoothly and $S_{\mu }^{\nu }=0$.

For our spacetime the components of this tensor which do not vanish
identically are equal to

\begin{equation}
K_{0}^{0}=-\frac{b^{\prime }}{b}\text{, }K_{2}^{2}=-\frac{r^{\prime }}{r}%
=K_{3}^{3}\text{, }K=-\frac{2r^{\prime }}{r}-\frac{b^{\prime }}{b}\text{,}
\end{equation}
where prime denotes differentiation with respect to the proper length $l$.
We have 
\begin{equation}
8\pi S_{2}^{2}=[K_{2}^{2}]-[K]=-[K_{0}^{0}]-[K_{2}^{2}]=\frac{%
(br)_{+}^{\prime }-(br)_{-}^{\prime }}{br}  \label{s22}
\end{equation}
\begin{equation}
8\pi S_{0}^{0}=[K_{0}^{0}]-[K]=-2[K_{2}^{2}]=\frac{2(r_{h}^{\prime
}-r_{-}^{\prime })}{r}\text{,}  \label{s00}
\end{equation}
\begin{equation}
\tilde{T}_{\mu }^{\nu }=S_{\mu }^{\nu }\delta (l-l_{0})\text{, }%
b_{0}=b(r_{0})\text{.}
\end{equation}

By assumption, the outer region is the RN one, the index ''0'' corresponds
to the shell. We consider the mixed components of tensors $S_{\mu }^{\nu }$
and $K_{\mu }^{\nu }$ with one upper and one lower indices since in the
metric (\ref{b}) they correspond to the orthonormal frame, so that on the
horizon, where the coordinate frame (\ref{b}) fails, they remain
well-defined.

Then for the ''+'' region the metric can be written in the curvature
coordinates like 
\begin{equation}
ds^{2}=-dt^{2}f+f^{-1}dr^{2}+r^{2}d\omega ^{2}\text{, }f=b^{2}\text{.}
\label{cur}
\end{equation}

\begin{equation}
b^{2}=1-\frac{2m}{r}+\frac{e^{2}}{r^{2}}\text{.}  \label{b2}
\end{equation}

One should distinguish two cases of the metric inside the shell.

\section{gluing Reissner-Nordstr\"{o}m and Minkowski metrics.}

First, following \cite{vf}, we consider the empty (Minkowski) spacetime
inside. Then in the \textquotedblright -\textquotedblright\ region $b=1$. In
the non-extremal case one obtains, in accordance with \cite{vf} that smooth
matching is impossible: on the horizon $S_{0}^{0}\rightarrow -\frac{1}{4\pi
r_{h}}\neq 0$, $S_{2}^{2}\rightarrow \frac{\sqrt{m^{2}-e^{2}}}{r_{h}^{2}b}%
\rightarrow \infty $. Much more interesting is the extremal case. Let now $%
m=e$. Then 
\begin{equation}
b=1-\frac{m}{r}\text{, }S_{2}^{2}=0=S_{3}^{3}\text{, }8\pi S_{0}^{0}=-\frac{%
2m}{r_{0}^{2}}\text{, }  \label{se}
\end{equation}

\begin{equation}
\lim_{r_{0}\rightarrow r_{h}}S_{0}^{0}=-\frac{1}{4\pi r_{h}}\neq 0\text{.}
\end{equation}
Our conclusion that bare stresses for the extremal case do not vanish
contradicts to the statement in \cite{vf}. To understand the source of
discrepancy, let us use, say, isotropic coordinates \cite{prep} whose
advantage consists in the continuity of the $g_{11}$ coefficient across the
shell at $\rho =a$:

\begin{equation}
ds^{2}=-b^{2}(\rho )dt^{2}+c^{2}(\rho )(d\rho ^{2}+\rho ^{2}d\omega
^{2})=-b^{2}(\rho )dt^{2}+c^{2}(\rho )(dx^{2}+dy^{2}+dz^{2})\text{, }r=\rho
c(\rho )
\end{equation}
for $\rho \geq a$ and 
\begin{equation}
ds^{2}=-b^{2}(a)dt^{2}+c^{2}(a)(d\rho ^{2}+\rho ^{2}d\omega ^{2})
\end{equation}
for $\rho \leq a$,

\begin{equation}
b=\frac{1}{c}\text{, }c=1+\frac{m}{\rho }\text{.}
\end{equation}

The tensor $\tilde{T}_{\nu }^{\mu }$ for the case under consideration looks
like 
\begin{equation}
\tilde{T}_{0}^{0}=B\delta (\rho -a)\text{, }B=-\frac{ma}{4\pi (m+a)^{3}}%
\text{,}  \label{t}
\end{equation}
all other components vanishing identically. If the shell approaches the
horizon, $a\rightarrow 0$ and $B\rightarrow 0$. It {\it would seem} that in
this limit matching becomes smooth. However, the point is that it is $%
S_{0}^{0}$ but not $\tilde{T}_{0}^{0}$ is the relevant quantity that
determines whether matching is smooth or not. Direct evaluation according to
(\ref{s}) gives non-zero $S_{0}^{0}$ (\ref{se}) due to the factor $c$ in the
proper distance $dl=cd\rho $ in the definition (\ref{s}). Moreover, the
coefficient $B$ (in contrast to $S_{0}^{0}$) is coordinate dependent. Using,
for example, the proper distance, one would obtain instead of $B$ the
quantity $\tilde{B}=Bc=S_{0}^{0}$ that does not vanish when $\rho
\rightarrow a$, $l\rightarrow \infty $. Thus, in the coordinate frame under
discussion $\lim_{r_{0}\rightarrow r_{h}}\tilde{T}_{\nu }^{\mu }=0$ but $%
\lim_{r_{0}\rightarrow r_{h}}S_{\nu }^{\mu }\neq 0$, the latter fact being
independent of the frame. (As far as the quantity $\tilde{T}_{0}^{0}\sqrt{-g}
$ used in \cite{vf} is concerned, it does vanish on the horizon even in the
case of an non-extremal black hole due to the addition factor $b.$ However,
it cannot be used as a criterion of smooth matching).

The impossibility of smooth matching can be also understood as follows. It
is shown in \cite{ms}, \cite{pod} that for any time-like surface $r=r_{0}(t)$
interior and exterior match smoothly, only if the condition 
\begin{equation}
m_{-}\equiv m(r_{0}-0)=m_{+}\equiv m(r_{0}+0)
\end{equation}
is satisfied, 
\begin{equation}
m(r)\equiv \frac{r}{2}[1-(\nabla r)^{2}]\text{.}  \label{mass}
\end{equation}

For the Minkowski metric $m_{-}=0$. For RN metric $m_{+}(r)=m-\frac{e^{2}}{2r%
}$. In particular, $m_{+}(\infty )=m$, $m_{+}(r_{h})=m-\frac{e^{2}}{2r_{h}}=%
\frac{2m^{2}-e^{2}+2m\sqrt{m^{2}-e^{2}}}{2(m+\sqrt{m^{2}-e^{2}})}$. Both in
the non-extremal and extremal cases $m_{+}\neq 0$. Thus, one cannot sew
smoothly both regions contrary to was stated in \cite{vf}, \cite{prep}.

The fact that $\lim_{r_{0}\rightarrow r_{h}}S_{0}^{0}\neq 0$ actually means
that the proper mass $m_{p}$ of the layer is finite. Indeed, 
\begin{equation}
m_{p}=4\pi \int_{r_{0}-0}^{r_{0}+0}dlr^{2}(-\tilde{T}_{0}^{0})=-4\pi
r_{h}^{2}S_{0}^{0}=r_{h}=m\neq 0\text{.}
\end{equation}

Thus, the {\it proper} mass of the layer is equal to the active mass.

On the other hand, the contribution of the layer to the active mass $m=4\pi
\int dr(-\tilde{T}_{0}^{0})r^{2}dr$ vanishes in the horizon limit due to an
additional factor $b(r_{0})\rightarrow 0$.

\section{Gluing Reissner-Nordstr\"{o}m and Bertotti-Robinson metrics}

We saw that the classical model of an elementary particle based on matching
RN and Minkowski metrics suffers from discontinuity in the geometry.
Meanwhile, there is another possibility due to the fact that in the case $%
m=e $ there is a special branch of solutions of field equations, apart from
the extremal RN metric. This is nothing else than the Bertotti-Robinson
metric (BR) \cite{r}, \cite{b} that is characterized by the property $%
r(l)=const$. In particular, as is well known, in the near-horizon region the
metric of the extremal RN tends to that of the BR. It also appears naturally
in the thermodynamic context as the extremal limit of non-extremal
configurations \cite{limit1}, \cite{limit2}. However, it does not entail
immediately that RN and BR metrics match smoothly in the limit under
discussion since the quantities like $K_{\mu }^{\nu }$ involve not only the
metric itself but also first derivatives. There exist different forms of the
BR metrics corresponding to the non-extremal ($b=\sinh \frac{l}{r_{0}}$) and
extremal versions ($b=\exp (l/r_{0})$) and also to the absence of the
horizon at all ($b=\cosh (l/r_{0})$). This is connected with the existence
of three independent Killing time-like vectors, the horizons being in the
case under discussion acceleration (not a black hole) horizons (see, for
details, e.g., \cite{lap}).

\subsection{The horizon limit of time-like shells}

It follows from the continuity of $g_{00}$ that the only suitable candidate
for matching is the extremal BR metric,

\begin{equation}
b=\exp (l/r_{0})\text{.}
\end{equation}

Then direct calculations give us 
\begin{equation}
S_{2}^{2}=S_{3}^{3}=0\text{, }S_{0}^{0}=\frac{b_{0}}{4\pi r_{0}}\text{. }
\end{equation}

In the limit $r_{0}\rightarrow r_{h}$ we have $S_{\mu }^{\nu }\rightarrow 0$%
. Moreover, the proper mass $m_{p}=4\pi
\int_{r_{0}-0}^{r_{0}+0}drr^{2}b^{-1}(-\tilde{T}_{0}^{0})=-4\pi
S_{0}^{0}r_{0}^{2}$ of the transition layer is negative and vanishes in this
limit: 
\begin{equation}
m_{p}=-b_{0}r_{0}\rightarrow 0\text{.}
\end{equation}

Again, the fact that both metrics match smoothly, can be understood in terms
of the effective mass (\ref{mass}). For the BR metric $r=r_{h}=const$, so $%
(\nabla r)^{2}\equiv 0$. For the extremal RN $(\nabla r)^{2}$ does not
vanish identically but tends to zero on the horizon. As $r_{0}\rightarrow
r_{h}$, masses coincide from both sides of the surface ($m_{-}=m_{+}=\frac{e%
}{2}$) and this makes smooth gluing possible.

Thus, for any $r_{0}\neq r_{h}$ it is impossible to glue smoothly RN and BR
spacetimes but, as $r_{0}$ approaches the horizon, mismatch becomes smaller
and smaller and disappears in the limit $r_{0}=r_{h}$.

\subsection{Light-like shells}

We discussed matching along the sequence of timelike surfaces. One may ask
the question, what happens to a light-like shell if one places it on the
horizon $r=r_{h}$ from the very beginning? In general, because of different
conditions of matching, one cannot expect the result to coincide with the
light-like limit of time-like shells. In particular, if a shell is placed
along the line $u=const$, where $u$ is an isotropic coordinate, only $S_{uu}$
can survive and $S_{2}^{2}$ vanish, whereas for time-like shells it remains
nonzero in the non-extremal case. Matching of two different non-extremal RN
black holes and non-extremal RN and Minkowski metric along light-like
surfaces was considered in \cite{dray} and it follows from the corresponding
results that stresses on the shell do not vanish. Now let us discuss the
case of the extremal RN. It is obvious that Minkowski spacetime cannot be
glued to RN along the horizon since the surface $r=r_{h}=const$ is not
ligh-like in the Minkowski metric. Instead, we again discuss the possibility
of smooth gluing between the extremal RN and BR spacetimes.

For the spherically-symmetrical case it is sufficient to use the condition
derived in \cite{ber} (the most general formalism for light-like shell is
developed in \cite{bi}). Let us write the metric in the form

\begin{equation}
ds^{2}=-H(U,V)dUdV+r^{2}d\Omega ^{2}\text{,}
\end{equation}
where $U$ and $V$ are the Kruskal-like coordinate in which the metric
coefficient $H$ remains bounded on the horizon. For definiteness, consider
the future horizon $U=0$.

The condition of matching along $U=0$ follows from eq. (6.14) of Ref. \cite%
{ber} and reads: 
\begin{equation}
\left( \frac{\partial r}{\partial U}\right) _{+}-\left( \frac{\partial r}{%
\partial U}\right) _{-}=0\text{.}  \label{ru}
\end{equation}

The explicit construction of the Kruskal-like coordinates for the extremal
case was carried out in \cite{lib}, where the coordinates $U$ and $V$ are
defined according to 
\begin{equation}
u=-\psi (-U)\text{, }v=\psi (V)\text{, }u=t-r_{*}\text{, }v=t+r_{*}\text{,}
\end{equation}
the tortoise coordinate

\begin{equation}
r_{*}=\int \frac{dr}{b^{2}}=r+\frac{1}{2}\psi (r-r_{h})\text{, }b=\left( 1-%
\frac{r_{h}}{r}\right) \text{, }\psi (\xi )\equiv 4r_{h}(\ln \xi -\frac{r_{h}%
}{2\xi })\text{.}
\end{equation}
In the vicinity of the future horizon $U=0$ it follows that $U=-(r-r_{h})$ 
\cite{lib}, so on the horizon $\left( \frac{\partial r}{\partial U}\right)
_{RN}=-1\neq \left( \frac{\partial r}{\partial U}\right) _{BR}=0$. Thus,
smooth matching is impossible.

\section{Energy associated with horizon}

\subsection{Acceleration horizons and gravitational mass-defect}

As it was shown in \cite{limit1}, \cite{limit2}, there exists such a
limiting transition from near-extremal black hole to the extremal state that
a black hole horizons turns into the acceleration one, typical of the BR
spacetime and, in doing so, all point of manifold pick up the value $r=r_{h}$%
. The similar conclusion is valid if the RN metric is extremal from the very
beginning. The mass between two values $r=r_{1}$ and $r=r_{2}$ $m(1,2)=4\pi
\int_{r_{1}}^{r_{2}}(-T_{0}^{0})r^{2}dr$. As $T_{0}^{0}$ is finite and $%
r_{1}\rightarrow r_{2}\rightarrow r_{h}=e$ in this limit, $m(1,2)\rightarrow
0$. On the other hand, the proper mass of the same region 
\begin{equation}
m_{p}=4\pi \int_{r_{1}}^{r_{2}}(-T_{0}^{0})r^{2}dl\rightarrow \frac{l(1,2)}{2%
}\text{,}
\end{equation}
where $l(1,2)$ is the proper distance between points $1$ an $2$. If one of
them corresponds to the horizon, $m_{p}\rightarrow \infty $ since $%
l\rightarrow \infty $. This is the feature inherent of the extremal horizon
independent of the concrete form of the metric and is valid, in particular,
for the RN and BR spacetimes.

On the other hand, the quasilocal energy \cite{by} 
\begin{equation}
E=4\pi r^{2}\varepsilon \text{, }\varepsilon =\frac{k-k_{0}}{8\pi }\text{, }%
k=-2b\text{,}
\end{equation}
for the flat spacetime $k_{0}=-2$. Here $k$ is the mean curvature of the
two-dimensional surface $r=r_{0}$ embedded into the three-metric$.$ It was
shown in \cite{can} that such an energy appears naturally in the
thermodynamic context for generic bounded self-gravitating static systems.
When $r_{0}\rightarrow r_{h}$, $k\rightarrow 0$ and $E=E_{0}=const$ does not
depend on $l$ (i.e. the position of the boundary).

Thus, the BR spacetime is an example of spacetimes which give the ultimate
case of the gravitational mass-defect: the active mass $m=const$, the energy 
$E=const$. The proper mass between a horizon and any other point is infinite
in the extremal case. For the non-extremal one it is finite but the total
amount integrated over all manifold is infinite. This situation is typical
of acceleration horizons (it is worth noting that for such horizons not only
the energy but also thermodynamics becomes in some sense degenerate \cite{ac}%
). These properties of the energy are similar to those of so-called T-models
which were analyzed carefully by Ruban \cite{rub1}, \cite{rub2}. In both
cases the coefficient at the angular part of the metric $r$ does not depend
on $l$ but T-models are actually cosmological, the time dependence of $r(t)$
being essential, whereas in the BR case $r=const$. Thus, acceleration
horizons give one more way of ultimate gravitation binding of an infinite
amount of energy into a finite active mass and quasilocal energy.

\subsection{VF model and M\o ller pseudotensorThus, acceleration horizons
give one more way of ultimate gravitation binding of an infinite amount of
energy into a finite active mass and quasilocal energy.}

It was one of the main statements in \cite{vf} that the extremal RN horizon
contributes nothing into dynamics. This conclusion was reached on the basis
of the M\o ller pseudotensor. Omitting details, the energy-momentum vector
of the system can be written as 
\begin{equation}
P_{\mu }=P_{\mu }^{\infty }-P_{\mu }^{h}\text{,}
\end{equation}
where $P_{\mu }^{\infty }=\frac{1}{2}\oint_{r\rightarrow \infty _{h}}U_{\mu
}^{\nu \sigma }d\sigma _{\nu \sigma }$ is calculated at infinity, $P_{\mu
}^{h}=\frac{1}{2}\oint_{r=r_{h}}U_{\mu }^{\nu \sigma }d\sigma _{\nu \sigma }$
is the contribution form horizon, $U_{\mu }^{\nu \sigma }$ is the
superpotential. The integration in $P_{\mu }^{h}$ is carried out over the
two-dimensional surface obtained as intersection of the horizon $r=r_{h}$
and some three-dimensional space-like surface depending on the foliation.

It is shown in \cite{vf} for the RN metric that (\ref{cur}) $%
U_{0}^{ik}=U_{i}^{0k}=0$ ($i=1,2,3$), 
\begin{equation}
U_{0}^{0k}=-\frac{x^{k}}{4\pi r^{2}}b(b-1)\text{.}
\end{equation}
\begin{equation}
U_{i}^{kl}=\frac{b}{8\pi }(\frac{b-1}{r}+b^{\prime })(\frac{x^{k}}{r}\delta
_{il}-\frac{x^{l}}{r}\delta _{ik})\text{,}
\end{equation}
$x^{k}$ are quasi-Cartesian coordinates related to $r$, $\theta $, $\phi $
in the same manner like in the usual flat space.

For the foliation of the spacetime by space-like hypersurfaces $t=const$,
the element of the two-dimensinal surface $d\sigma _{\mu \nu }$ has
non-vanishing components $d\sigma _{0k}$ only, the term $U_{0}^{0k}d\sigma
_{0k}$ vanishes on the horizon due to the factor $b$, so that $P_{\mu
}^{h}=0 $. However, if one uses some other foliation with another space-like
surfaces, the terms with $U_{i}^{kl}$ give rise, in general, to $P_{\mu
}^{h}\neq 0$. To make it vanish and, thus, to achieve the zero contribution
from the horizon independent of foliation, Vilenkin and Fomin demand that $%
U_{0}^{kl}(r_{h})=0$, whence $\left( b^{2}\right) _{r=r_{h}}^{\prime }=0$.
This entails that the black hole should be extremal, $m=e$. As the quantity $%
U_{\mu }^{\nu \sigma }d\sigma _{\nu \sigma }$ is a vector (since both
factors are tensor densities of opposite weights), it is concluded in \cite%
{vf} that for the extremal case the equality $Q_{\mu }\equiv U_{\mu }^{\nu
\sigma }d\sigma _{\nu \sigma }=0$ holds in any coordinate system.

However, the fact that the covariant components of the vector vanish in the
system which itself is ill-defined on the horizon, should not entail the
conclusion about vanishing this vector as such. To clarify the essence of
matter, let us consider a more general situation when some vector $Q_{\mu }$
has in the coordinate system (\ref{cur}) the components $Q_{\mu }=(Q_{0},0)$%
, where $Q_{0}=bb^{\prime }A$, $A\neq 0$ is finite on the horizon. Let us
consider at first the non-extremal case, when $b\sim \sqrt{r-r_{h}}$. It 
{\it would seem} that, as $Q_{0}\rightarrow 0$ as $r\rightarrow r_{h}$, the
vector $Q_{\mu }(r_{h})=0$ and, moreover, this equality holds in any
coordinate system since $Q_{\mu }$ is a vector. However, it is easy to see
that the vector norm $Q_{\mu }Q^{\mu }=g^{00}Q_{0}^{2}$ remains non-zero on
the horizon because of the factor $g^{00}\sim (r-r_{h})^{-1}$. In the
Kruskal-like coordinates $U$, $V$ in which the metric coefficients are
well-defined, direct check shows that $Q_{U}$, $Q_{V}\neq 0$ near the point $%
U=0=V$ that corresponds to the surface $t=const$. For the extremal horizon $%
Q_{0}\sim r-r_{h}$ but $g^{00}\sim (r-r_{h})^{-2}$ and, again, $Q_{\mu
}Q^{\mu }\neq 0$ on the horizon.

The fact that in the static frame $Q_{\mu }(r_{h})=0$ can be indeed
interpreted as a manifestation of freezing dynamics but from the viewpoint
of an external observer only. Correspondingly, the conclusion that for the
extremal case $P_{\mu }$ can be defined in the outer region only, with the
contribution from the horizon vanishing \cite{vf}, retains its validity in
such a frame. However, another observer, who is diving inside a black hole,
will find that the horizon does contribute into the dynamics of the system.

\subsection{Quasilocal energy of horizon}

In the modern approach, there is no necessity to resort to pseudotensors for
constructing dynamic characteristics of gravitational field. Quasilocal
energy and momentum are defined on the basis of the action principle \cite%
{by}, \cite{bly}. A reader can address these papers for a detailed
formalism, here we only borrow from there some general results. Consider the
spacetime region $M$ with the boundary $\partial M$ that consists of a
timelike element $\bar{T}$ and spacelike elements $\Sigma ^{\prime }$ and $%
\Sigma ^{\prime \prime }$ which are leaves of foliation defined by $t=const$%
. The intersections of $\Sigma $ leaves with $\bar{T}$ define a foliation of 
$\bar{T}$ into two-dimensional spacelike surfaces $B$.

Let $u_{\mu }=-Nt_{,\mu }$ be the unit four-velocity for a family of $\Sigma 
$ leaves, the lapse function $N$ ensures the condition $u_{\mu }u^{\mu }=-1$%
. Let us also consider the foliation of the spacetime $M$ by the family of
timelike surfaces $s=const$ that contains $\bar{T}$ as one of its leaves and
introduce the unit-normal vector $\bar{n}_{\mu }=\bar{M}\nabla _{\mu }s$,
where $\bar{M}$ ensures normalization $\bar{n}^{\mu }\bar{n}_{\mu }=1$. Also
define the unit vector $\bar{u}_{\mu }=-\bar{N}D_{\mu }t$, where $D_{\mu }$
is the covariant derivative on $s=const$. Then, one can define the
quasilocal densities for the energy $\varepsilon $, normal momentum $%
j_{\vdash }$, tangential momentum and temporal stresses. We focus our
attention on the first two quantities and the law of their transformation.
There are two sets of observers connected by local boosts with the relative
velocity $v$. The first one consists of those comoving with $\bar{T}$ and at
rest with respect to the $B$ foliation of $\bar{T}$ (characterized by the
barred quantities), while the second one consists of those at rest with
respect to the $\Sigma $ foliation (characterized by the unbarred
quantities).

The energy density and normal momentum $j_{\vdash }$ are equal to 
\begin{equation}
\kappa \varepsilon =k\text{, }\kappa j_{\vdash }=-\sigma ^{ij}K_{ij},\kappa
=8\pi \text{,}  \label{d}
\end{equation}
where $K_{ij}$ is the extrinsic curvature tensor associated with the
spacelike hypersurface $\Sigma $, $k$ is the mean curvature of the
two-dimensional boundary $^{2}B$ with the metric $\sigma _{ab}$ embedded in $%
\Sigma $. Similar quantities are defined for barred observers with the
transformation law under local boosts \cite{bly} 
\begin{equation}
\bar{\varepsilon}=\gamma \varepsilon -\gamma vj_{\vdash }\text{, }\bar{j}%
_{\vdash }=\gamma j_{\vdash }-\gamma v\varepsilon \text{,}  \label{l1}
\end{equation}
\begin{equation}
\varepsilon =\gamma \bar{\varepsilon}+\gamma v\text{ }\bar{j}_{\vdash }\text{%
, }j_{\vdash }=\gamma \bar{j}_{\vdash }+\gamma v\varepsilon \text{,}
\label{l2}
\end{equation}

where 
\begin{equation}
\gamma =-u_{\mu }\bar{u}^{\mu }=(1-v^{2})^{-1/2}  \label{gamma}
\end{equation}
is the Lorentz factor.

Let the RN metric be written in the form

\begin{equation}
ds^{2}=-N^{2}dt^{2}+H^{2}dr^{2}+R^{2}d\Omega ^{2}\text{,}  \label{nh}
\end{equation}
where all metric coefficients, which depend in general on $r$ and $t$, are
regular on the horizon (coordinates of the Graves and Brill type \cite{gb}).
Our goal is to compare dynamic characteristics of the surface $r=r_{0}$ in
the limit $r_{0}\rightarrow r_{h}$ for two sets of aforementioned observers.
We use the barred quantities for the observers comoving with respect to the
boundary element $\bar{T}$ that in our case represents the surface $%
R-R_{0}=0 $, $R_{0}=const$. Then in the coordinates (\ref{nh}) 
\begin{equation}
\bar{n}_{\mu }=\alpha (\dot{R},R^{\prime },0,0)\text{, }\bar{u}_{\mu
}=\alpha (-R^{\prime }\frac{N}{H},-\dot{R}\frac{H}{N},0,0)\text{, }\alpha
^{-2}\equiv (\nabla R)^{2}=\frac{R^{\prime 2}}{H^{2}}-\frac{\dot{R}^{2}}{%
N^{2}}\text{,}
\end{equation}
\begin{equation}
\bar{n}_{\mu }=\alpha (\dot{R},R^{\prime },0,0)\text{, }\bar{u}_{\mu
}=\alpha (-R^{\prime }\frac{N}{H},-\dot{R}\frac{H}{N},0,0)\text{, }\alpha
^{-2}\equiv (\nabla R)^{2}=\frac{R^{\prime 2}}{H^{2}}-\frac{\dot{R}^{2}}{%
N^{2}}\text{,}  \label{u}
\end{equation}
where $\bar{n}_{\mu }\bar{u}^{\mu }=0$, $\bar{n}_{\mu }$ is a spacelike unit
vector. In the curvature coordinates (\ref{cur}) our surface looks like $%
r=const$ and the same vector $\bar{u}_{\mu }$ written in these coordinates
has the typical form $-\sqrt{f}(1,0,0,0)$.

We also consider observers who are at rest with respect to the slices of the
constant Graves and Brill time $t$ (\ref{nh}) but move from the viewpoint of
observers who use static coordinates (\ref{cur}). For such observers, using
notations without bar, we have in the coordinates (\ref{nh}): 
\begin{equation}
u_{\mu }=-N(1,0,0,0)\text{, }n^{\mu }=H^{-1}(0,1,0,0)\text{. }  \label{u1}
\end{equation}

We can calculate the energy density $\varepsilon $ by two methods - directly
from (\ref{d}) or on the basis of the transformation law (\ref{l2}).

For the foliation (\ref{u1})

\begin{equation}
K_{\theta }^{\theta }=K_{\phi }^{\phi }=-\frac{\dot{R}}{RN}\text{,}
\end{equation}
\begin{equation}
\kappa \varepsilon =-\frac{2R^{\prime }}{HR}\text{.}  \label{ek}
\end{equation}
\begin{equation}
\kappa j_{\vdash }=-2\frac{\dot{R}}{RN}\text{.}
\end{equation}

To find $\bar{\varepsilon}$, it is convenient to use the coordinates (\ref%
{cur}). Then $\bar{j}_{\vdash }=0$, 
\begin{equation}
\kappa \bar{\varepsilon}=-k=-2\frac{\sqrt{f}}{R}\text{.}
\end{equation}

It follows from (\ref{gamma}), (\ref{u}), (\ref{u1}) that 
\begin{equation}
\gamma =\frac{R^{\prime }}{H\left| \nabla R\right| }\text{.}
\end{equation}

Expressing the scalar $\left| \nabla R\right| \equiv \sqrt{(\nabla R)^{2}}$
in curvature coordinates, where $R=r$, we see that $\left| \nabla R\right| =%
\sqrt{f}$. Now we may exploit the formula (\ref{l2}), where only radial
boosts are relevant which do not touch upon angle variable $\theta $ and $%
\phi $. We have 
\begin{equation}
\varepsilon =\gamma \bar{\varepsilon}=-\frac{2R^{\prime }}{HR}  \label{cure}
\end{equation}
that again leads to (\ref{ek}). We see that (\ref{ek}) agrees with (\ref%
{cure}) and, thus, both methods of calculations (for unbarred quantities at
once or for barred with the subsequent boost) give the same result. It is
worth stressing that, as one approaches the horizon, $f\rightarrow 0$, $\bar{%
\varepsilon}\rightarrow 0$ but $\gamma \rightarrow \infty $, so that the
product $\varepsilon =\gamma \bar{\varepsilon}$ remains finite. It is worth
noting that the fact that $v\rightarrow 1$, $\gamma \rightarrow \infty $
means that the static system (curvature coordinates) becomes ill-defined:
its relative speed to comoving observers approaches that of light.

It is also instructive to calculate the invariant $M^{2}\equiv \left( \kappa
\varepsilon \right) ^{2}-\left( \kappa j_{\vdash }\right) ^{2}=\kappa
^{2}p_{\mu }p^{\mu }$, where $p_{\mu }=(\varepsilon $, $j_{\vdash })$ \cite%
{bly}. Then one obtains

\begin{equation}
M^{2}=\frac{4}{R^{2}}(\frac{R^{\prime 2}}{H^{2}}-\frac{\dot{R}^{2}}{N^{2}})=%
\frac{4}{R^{2}}\left( \nabla R\right) ^{2}\text{.}  \label{inv}
\end{equation}

Thus, from the fact that near the horizon $\bar{\varepsilon}\rightarrow 0$
and $\bar{j}_{\vdash }=0$ in static coordinates it does {\it not }follow
that the vector $p_{\mu }$ vanishes in any frame, be the horizon an extremal
or non-extremal. Rather, this vector becomes isotropic on the horizon. To
probe it, an observer should use the Graves and Brill reference frame, in
other words he should fall into a black hole.

As by-product, we see from (\ref{inv}) that for spherically-symmetrical
generic spacetimes $M^{2}>0$ in so-called R-region, $M^{2}<0$ in T-regions 
\cite{novikov} and $M^{2}=0$ on the horizons or in regions where $\left(
\nabla R\right) ^{2}$ is isotropic. For the BR spacetimes, when $R\equiv
const$, $M^{2}=0$ as well. It is also seen from (\ref{mass}) that $m=\frac{R%
}{2}(1-\frac{M^{2}R^{2}}{4})$.

\section{Extremal Kerr-Newman geometry and rotating analog of
Bertotti-Robinson spacetime}

In this section we generalize the results typical of RN metric to the case
of the Kerr-Newman (KN) one. Namely, we consider the extremal KN metric and
sew it with the rotating analog of the BR. (There is no sense to consider
also the non-extremal version since even in the non-rotational case matching
under discussion is impossible, as is shown in previous sections.) Next, we
show that in the horizon limit this matching becomes smooth. The extremal KN
metric has the general form ($x^{0}=t$, $x^{1}=r$, $x^{2}=\theta $, $%
x^{3}=\phi $)

\begin{equation}
ds^{2}=-N^{2}dt^{2}+A^{2}dr^{2}+\rho ^{2}d\theta ^{2}+D^{2}(d\phi +Vdt)^{2}%
\text{,}  \label{kn1}
\end{equation}
\begin{equation}
A^{2}\equiv B^{^{-}2}\text{, }B=\frac{(r-r_{h})}{\rho }\text{, }\rho
^{2}=r^{2}+a^{2}\cos ^{2}\theta \text{, }  \label{kn2}
\end{equation}
all metric coefficients do not depend on $t$ and $\phi $, the coefficient $D$
(whose explicit form is irrelevant for us) is finite on the horizon.

The lapse and shift functions 
\begin{equation}
V^{\phi }\equiv V=(r-r_{h})q(x^{1},x^{2})\text{, }N=(r-r_{h})\chi
(x^{1},x^{2})\text{, }  \label{vn}
\end{equation}
where in the vicinity of the horizon $q$ and $\chi $ are finite. Our frame
corotates with the horizon, so that on the horizon $V\rightarrow 0$ and $%
\frac{V^{\phi }}{N}\rightarrow \frac{q_{+}}{\chi _{+}}$, where $f_{+}$ means 
$f(r_{h}$, $\theta )$.

One can obtain from the KN the rotating analog of the BR metric (RBR) just
as BR can be obtained from the near-extremal or extremal RN. For the
non-extremal case such a procedure was carried out in \cite{hor} and for the
extremal case in \cite{bar}. As we are dealing with the extremal horizon,
the limiting transition of \cite{bar} is now relevant. Making the coordinate
transformation 
\begin{equation}
r=r_{h}+\lambda \tilde{r}\text{, }t=\frac{\tilde{t}}{\lambda }\text{,}
\end{equation}
one obtains in the limit $\lambda \rightarrow 0$ the extremal version of RBR
in the form 
\begin{equation}
ds^{2}=-\chi _{+}^{2}(x^{i})\tilde{r}^{2}d\tilde{t}^{2}+\frac{d\tilde{r}^{2}%
}{\tilde{r}^{2}}\rho _{+}^{2}+\rho _{+}^{2}d\theta ^{2}+D_{+}^{2}(d\phi +%
\tilde{V}d\tilde{t})^{2}\text{,}  \label{rbr}
\end{equation}
$\tilde{V}=q_{+}\tilde{r}$, $\frac{\tilde{V}}{\tilde{N}}=\frac{q_{+}}{\chi
_{+}}$, where $\chi _{+}$ and $q_{+}$ do not depend on $\tilde{r}$.

Let us consider the surface $r=r_{0}$ such that for $r>r_{0}$ the metric is
the KN and for $r<r_{0}$ it is the RBR, calculate $K_{\mu }^{\nu }$ from
both sides and compare the results. The unit vector orthogonal to the
surface has the components $n_{\mu }=(0,B,0,0)$. The extrinsic curvature
tensor reads 
\begin{equation}
K_{\iota j}=-n_{i;j},  \label{kij}
\end{equation}
where $i,j=0,2,3$ and the covariant derivative is calculated on the
hypersurface $r=r_{0}.$ We must consider the limiting transition $%
r_{0}\rightarrow r_{h}$ and compare the extrinsic curvature tensor for the
extremal KN metric with that for the RBR one. On the horizon the coordinate
frame (\ref{kn1}) - (\ref{vn}) becomes ill-defined but we overcome this
difficulty by using the orhonormal frame with basic vectors $h_{(a)}^{\mu }$%
, $a=0,1,2,3$ and calculating $K_{(a)(b)}=K_{\mu \nu }h_{(a)}^{\mu
}h_{(b)}^{(\nu )}$. It is convenient to choose the standard basic \cite{s}
(Ch. 2, Sec. 11) 
\begin{equation}
h_{(0)}^{\mu }=\frac{1}{N}(1,0,0,-V)\text{,}  \label{h0}
\end{equation}
\begin{equation}
h_{(1)}^{\mu }=B(0,1,0,0)\text{,}
\end{equation}
\begin{equation}
h_{(2)}^{\mu }=\frac{1}{\rho }(0,0,1,0)\text{, }
\end{equation}
\begin{equation}
h_{(3)}^{\mu }=\frac{1}{D}(0,0,0,1)\text{.}  \label{h3}
\end{equation}
Its typical feature consists in that it corresponds to local observers with
the zero angular momentum: the point with the four-velocity $u^{\mu }$ which
has the angular velocity $\omega =-V$ in the cordinate frame (\ref{kn1}) is
at rest in the locally inertial frame, so that $u^{(3)}=u_{(3)}=0$. In the
coordinate basic the formula (\ref{kij}) gives us 
\begin{equation}
K_{ik}=-\frac{B}{2}\frac{\partial g_{ik}}{\partial r}\text{. }  \label{kc}
\end{equation}
Then it follows from (\ref{h0}) - (\ref{h3}), (\ref{kc}) that (prime here
denotes differentiation with respect to $r$) for the KN metric 
\begin{equation}
K_{(2)(2)}^{KN}=-B\frac{\rho ^{\prime }}{\rho }\text{,}
\end{equation}
\begin{equation}
K_{(3)(3)}^{KN}=-B\frac{D^{\prime }}{D}\text{,}
\end{equation}
\begin{equation}
K_{(0)(0)}=B(\frac{N^{\prime }}{N}+2\frac{V^{2}DD^{\prime }}{N^{2}})\text{,}
\end{equation}
\begin{equation}
K_{(0)(3)}^{KN}=-\frac{V^{\prime }DB}{2N}\text{.}
\end{equation}
When $r_{0}\rightarrow r_{h}$, the coefficient $B\rightarrow 0$ and, using
the properties (\ref{kn2}), (\ref{vn}), we obtain $K_{(2)(2)}^{KN}%
\rightarrow 0$, $K_{(3)(3)}^{KN}\rightarrow 0$, $K_{(0)(3)}^{KN}\rightarrow -%
\frac{q_{+}D_{+}}{2\rho _{+}\chi _{+}}$, $K_{(0)(0)}\rightarrow \frac{1}{%
\rho _{+}}$.

From the other hand, if we calculate the extrinsic tensor for the RBR metric
(\ref{rbr}) we obtain 
\begin{equation}
K_{(2)(2)}^{RBR}=0=K_{(3)(3)}^{RBR}\text{ , }K_{(0)(3)}^{RBR}=-\frac{%
q_{+}D_{+}}{2\rho _{+}\chi _{+}}\text{, }K_{(0)(0)}^{RBR}=\frac{1}{\rho _{+}}%
\text{.}
\end{equation}%
We see that for all components 
\begin{equation}
\lim_{r_{0}\rightarrow r_{h}}K_{(a)(b)}^{KN}=K_{(a)(b)}^{RBR}.
\end{equation}

Thus, in the horizon limit the extremal KN geometry goes smoothly to that of
RBR spacetime.

\section{Summary and conclusions}

Pure field-theoretical models with bare stresses, vanishing in the horizon
limit, proved to be realized by means of sewing the extremal RN metric with
not the Minkowski, but rather BR spacetimes. Thus, the whole spacetime
reveals non-uniform topology structure: along with spherically-symmetrical
external part, it contains also direct product of two subspaces of constant
curvature inside. Our classical model of an elementary particle reveals
itself for an external observer as an extremal black hole, whose horizon is
situated at infinite proper distance. According to the properties of the BR
spacetime \cite{lap}, it extends infinitely without hitting a singularity
also beyond the horizon for an observer who dares to dive into it. Both for
the non-extremal or extremal cases the region inside the horizon does in
general contribute to dynamic characteristics. The energy-momentum vector,
associated with the horizon, turns out to be isotropic but non-vanishing. It
is of interest to generalize this result to generic isolated horizons \cite%
{ash1}.

We saw that for a family of time-like surfaces $r=r_{0}$ the magnitude of
the Lanczos tensor tends to zero when $r_{0}\rightarrow r_{h}$. On the other
hand, if $r_{0}=r_{h}$ exactly, the surface becomes light-like in which case
the matching conditions are qualitatively different and cannot be satisfied
for the case under discussion, when the BR metric is glued to the extremal
RN one. Thus, our construction should be understood in the sense of the
limiting transition only, in which case it gives ''mass without mass'' and
''charge without charge'' \cite{wh}. The singular residue for any member of
the family of configurations (before the limit is taken) can be understood
as follows. The applicability of the theorem about singularities inside
black holes implies that $\varepsilon \geq 0$ (weak energy condition) and $%
\varepsilon +\sum_{j}p_{j}\geq 0$ \cite{he} (Ch. 8.2). These conditions
break down, for example, for black holes with a dS core when $p=-\varepsilon
<0$ and, thus, regularity is achieved \cite{dym1}, \cite{dym2}. However,
they are satisfied for an electromagnetic field ($\varepsilon =w$, $p_{r}=-w$%
, $p_{\perp }=w$, $w=\frac{e^{2}}{r^{4}}$). Thus, had we had an everywhere
regular black hole metric, this would have contradicted the singularity
theorem. Our construction occupies an intermediate place between singular
and regular models: the ideal purely field classical Abraham - Lorentz model
of the electron remains unattainable but one may approach it as nearly as
one likes.

Our consideration was pure classical. As far as the role of quantum
backreaction is concerned, it was checked directly that the backreaction of
quantum massive fields changes the condition of extremality in such a way
that $m\neq e$ but does not prevent the existence of extremal horizons as
such, it also leaves intact the general geometrical character of
acceleration horizons as direct product of two spheres of constant curvature
(but, in contrast to the BR metric, their radii in the quantum corrected
case do not longer coincide) \cite{mz}. Therefore, one can glue the
quantum-corrected extremal RN metric to the quantum-corrected analog of the
BR one in the much the same way as it was done above classically. Bearing
also in mind that extremal horizons have the Hawking temperature $T_{H}=0$
and do not radiate, one can expect our static self-supported solutions as a
whole obtained pure classically survive also on the semiclassical level.

It is common belief that extremal black holes can be suitable candidates on
the role of remnants after black hole evaporation. Investigations of some
two-dimensional exactly solvable models with account for quantum
backreaction showed that remnants can also represent semi-infinite throats
corresponding to two-dimensional AdS spacetimes which are noting else than
two-dimensional analog of BR ({\rm AdS}$_{2}\times S^{2}$) \cite{bose}. One
is led to think that, perhaps, BR spacetime can also be relevant in this
context for late stages of evaporation of near-extremal black holes.

\section{Acknowledgment}

I am grateful to Alexander Vilenkin for suggesting the problem to me, which,
many years later, evoked a quite unexpected response.




%
%

%
%

\end{document}